\documentclass[twocolumn,showpacs]{revtex4}

\usepackage{epsfig}\makeatletter
\usepackage{graphicx}
\usepackage{dcolumn}\DeclareRobustCommand\bblash{\btt{\@backslashchar}}%
\usepackage{amsmath}\makeatother

\begin{document}

\title{Phantom Field with O(\textit{N}) symmetry in Exponential Potential}

\author{Xin-zhou Li}\email{kychz@shtu.edu.cn}
\author{Jian-gang Hao}
\affiliation{ Shanghai United Center for Astrophysics (SUCA),
Shanghai Normal University, Shanghai 200234, China}%

\date{\today}

\begin{abstract}
In this paper, we study the phase space of phantom model with
O(\emph{N}) symmetry in exponential potential. Different from the
model without O(\emph{N}) symmetry, the introduction of the
symmetry leads to a lower bound $w>-3$ on the equation of state
for the existence of stable phantom dominated attractor phase. The
reconstruction relation between the potential of O(\textit{N})
phantom system and red shift has been derived.
\end{abstract}

 \maketitle

\section{Introduction}

Astronomical measurements from Supernovae \cite{riess, perlmutter,
tonry} and CMB anisotropy \cite{bennett,Netterfield,Halverson}
independently confirm that about two-thirds of the energy density
in our Universe is dark energy whose effective equation of state
may lies in the range $-1.38<w<-0.82$\cite{melchiorri}. A recent
analysis for the new high red shift data\cite{pad} as well as dark
energy and dark matter halos\cite{kuhlen} also indicate that dark
energy with $w<-1$ seems more favorable. While the equation of
state of conventional quintessence models\cite{steinhardt}that
based on a scalar field and positive kinetic energy can not evolve
to the the regime of $w<-1$.  Some
authors\cite{caldwell1,sahni,parker,chiba,boisseau,schulz,faraoni,maor,onemli,
torres,carroll,frampton,hao,liu,caldwell2,gibbons,feinstein,sami,Nojiri,Dabrowski,meng,Johri,Cline,Lu,
Gonzalez-Diaz,Stefancic,Calcagni,Aguirregabiria,Jarv} investigated
phantom models which have negative kinetic energy and can realize
the $w<-1$ in its evolution. Although the introduction of a
phantom field causes many theoretical problems such as the
violation of some widely accepted energy condition and the rapid
vacuum decay\cite{carroll}, it is still very interesting in term
of fitting current observations. Comparing with the other
approaches to realizing the $w<-1$, such as the modification of
Friedman equation, it seems economical.

On the other hand, the role of complex scalar field as
quintessence has been studied in\cite{complex} and its
generalization to the fields with O(\textit{N}) symmetry was done
in Ref. \cite{li}. In this paper, we study the case that when
phantom models have an O(\textit{N}) symmetry constraint via a
specific model in which the potential is exponentially dependent
on the field. We show that the dynamical system admits only the
phantom dominated attractor phase and the introduction of
O(\textit{N}) symmetry leads to a lower bound on the equation of
state as $w>-3$, which is absent in the singlet scalar field
phantom models. It is worth noting that although the O(\textit{N})
phantom and the O(\textit{N}) quintessence lead to similar
equations, but they have very different dynamical evolution and
physical implications, especially they will lead to different
evolution of the equation of state $w$. We also derive the
relation between the red-shift and the potential of the scalar
fields, known as reconstruction relation. It turns out that the
potential relates to the red-shift the same way as that in
ordinary scalar fields theory while the variation of the fields
does in a quite different manner.

\section{ O(N) Phantom}

We start from the flat Robertson-Walker metric

\begin{equation}\label{metric}
ds^{2}=dt^{2}-a^{2}(t)d\textbf{x}^2
\end{equation}

\noindent The Lagrangian density for the Phantom with
 O(\textit{N}) symmetry is

\begin{equation}
L_{\Phi}=-\frac{1}{2}g^{\mu\nu}(\partial_{\mu}\Phi^{a})(\partial_{\nu}\Phi^{a})-V(|\Phi^{a}|)
\end{equation}

\noindent where $\Phi^{a}$ is the component of the scalar field,
$a=1,2,\cdots,N$. To make it possess a O(\textit{N}) symmetry, we
write it in the following form

\begin{eqnarray}\label{imag}
\Phi^{1}=R(t)\cos\varphi_{1}(t)\hspace{4.2cm}\nonumber\\
\Phi^{2}=R(t)\sin\varphi_{1}(t)\cos\varphi_{2}(t)\hspace{2.85cm}\nonumber\\
\Phi^{3}=R(t)\sin\varphi_{1}(t)\sin\varphi_{2}(t)\cos\varphi_{3}(t)\hspace{1.5cm}\\
\cdots\cdots\hspace{4cm}\nonumber\\
\Phi^{N-1}=R(t)\sin\varphi_{1}(t)\cdots\sin\varphi_{N-2}(t)\cos\varphi_{N-1}(t)\hspace{-0.35cm}\nonumber\\
\Phi^{N}=R(t)\sin\varphi_{1}(t)\cdots\sin\varphi_{N-2}(t)\sin\varphi_{N-1}(t)\nonumber
\end{eqnarray}

\noindent Therefore, we have $|\Phi^{a}|=R$ and assume that the
potential of the O(\textit{N}) phantom depends only on \textit{R}.

The action for the universe is :

\begin{equation}\label{action}
S=\int d^{4}x\sqrt{-g}(-\frac{1}{16\pi
G}R_{s}-p_{\gamma}+L_{\Phi})
\end{equation}

\noindent where $g$ is the determinant of the metric tensor $
g_{\mu\nu}$ , $G$ is the Newton's constant, $R_{s}$ is the Ricci
scalar, and $p_{\gamma}$ is the pressure of the baryotropic fluid
whose equation of state is $p_{\gamma}=(\gamma-1)\rho_{\gamma}$.
Varying the action, one can obtain the Einstein equations and the
radial equation of scalar fields as

\begin{eqnarray}
H^{2}&=&\frac{8\pi G}{3}[\rho_{\gamma}+\rho_{\Phi}]\\\nonumber
(\frac{\ddot{a}}{a})&=&-\frac{8\pi
G}{3}[(\frac{3\gamma}{2}-1)\rho_{\gamma}+2
p_{\Phi}+V(R)]\\\nonumber
\ddot{R}&+&3H\dot{R}-\frac{\Omega^{2}}{a^{6}R^{3}}-\frac{\partial
V(R)}{\partial R}=0
\end{eqnarray}

\noindent with
$\rho_{\Phi}=-\frac{1}{2}(\dot{R}^{2}+\frac{\Omega^{2}}{a^{6}R^{2}})+V(R)$
and
$p_{\Phi}=-\frac{1}{2}(\dot{R}^{2}+\frac{\Omega^{2}}{a^{6}R^{2}})-V(R)$
are the energy density and pressure of the $\Phi$ field
respectively, and $H$ is Hubble parameter. To obtain the above
equations, we make use of the fact that the "angular components"
of the system could be simplified by the first integrals and its
net contribution to the dynamics could be manifested by the term
containing $\Omega$ in the radial equation. For details of this,
one can refer to Ref.\cite{li}. The equation of state for the
O(\textit{N}) phantom is

\begin{equation}
w =\frac{p_\Phi}{\rho_\Phi}=
\frac{\dot{R}^{2}+\frac{\Omega^{2}}{a^{6}R^{2}}+2V(R)}{\dot{R}^{2}+
\frac{\Omega^{2}}{a^{6}R^{2}}-2V(R)}
\end{equation}

It is clear that the O(\textit{N}) phantom could realize the
equation of state $w<-1$ which is equivalent to

\begin{equation}
0<\dot{R}^{2}+\frac{\Omega^{2}}{a^{6}R^{2}}<2V(R)
\end{equation}

\noindent where the term$\frac{\Omega^{2}}{a^{6}R^{2}}$ comes from
the "total angular motion".

\section{ Attractor Property of O(\textit{N}) Phantom
}

In this section, we firstly investigate the attractor property of
the O(\textit{N}) phantom field in an exponential potential. To do
so, we rewrite the equations of motion as

\begin{equation}\label{sys1}
\dot{H}=-\frac{\kappa^2}{2}(\rho_\gamma+p_\gamma-\dot{R}^2 -
\frac{\Omega^2}{a^6R^2})
\end{equation}

\begin{equation}\label{sys2}
\dot{\rho_\gamma}=-3H(\rho_\gamma+p_\gamma)
\end{equation}

\begin{equation}\label{sys3}
\ddot{R}+3H\dot{R}-\frac{\Omega^2}{a^6R^3}+\lambda\kappa V(R)=0
\end{equation}

\begin{equation}\label{sys4}
H^2=\frac{\kappa^2}{3}[\rho_\gamma-\frac{1}{2}(\dot{R}^2+\frac{\Omega^2}{a^6R^2})+V(R)]
\end{equation}

The potential $V(R)$ here is specified as
$V(R)=V_0\exp(-\lambda\kappa R)$. Now, we will introduce the
following variables to obtain the autonomous system for the above
dynamical system. The variables could be defined as
$x=\frac{\kappa}{\sqrt{6}H}\dot{R}$,
$y=\frac{\kappa\sqrt{V(R)}}{\sqrt{3}H}$,
$z=\frac{\kappa}{\sqrt{6}H}\frac{\Omega}{a^3R}$,
$\xi=\frac{1}{\kappa R}$, and $N=\log a$. The equation
system(\ref{sys1})-(\ref{sys4}) become the following autonomous
system:

\begin{eqnarray}\label{auto}
\frac{dx}{dN}&&=\frac{3}{2}x[\gamma(1+x^2-y^2+z^2)-2(x^2+z^2)]\\\nonumber&&-(3x-\sqrt{6}z^2\xi+\sqrt{\frac{3}{2}}\lambda
y^2)\\\nonumber
\frac{dy}{dN}&&=\frac{3}{2}y[\gamma(1+x^2-y^2+z^2)-2(x^2+z^2)]-\sqrt{\frac{3}{2}}\lambda
xy\\\nonumber
\frac{dz}{dN}&&=-3z+\frac{3}{2}z[\gamma(1+x^2-y^2+z^2)-2(x^2+z^2)]-\sqrt{6}xz\xi\\\nonumber
\frac{d\xi}{dN}&&=-\sqrt{6}\xi^2x
\end{eqnarray}

Also, we have a constraint equation

\begin{equation}\label{constraint}
\Omega_R+\frac{\kappa^2\rho_{\gamma}}{3H^2}=1
\end{equation}

\noindent where
\begin{equation}\label{omegar}
\Omega_R=\frac{\kappa^2\rho_R}{3H^2}=y^2-x^2-z^2
\end{equation}

The equation of state for the O(\textit{N}) phantom could be
expressed in term of the new variables as

\begin{equation}\label{equaofstate}
 w_{\Phi}=\frac{p_{\Phi}}{\rho_{\Phi}}=\frac{x^2+y^2+z^2}{x^2-y^2+z^2}
\end{equation}

The critical points of the above autonomous system are easily
obtained by setting the right hand sides of the above equations to
zero. We write the variables near the critical points $(x_c, y_c,
z_c, \xi_c)$ in the form $x=x_c+u$, $y=y_c+v$, $z=z_c+w$ and
$\xi=\xi_c+\chi$, where $u,v,w,\chi$ are perturbations of the
variables near the critical points and form a column vector
denoted as $\textbf{U}$ . Substitute the above expression into the
autonomous system (\ref{auto}), one can obtain the equations for
the perturbations up to the first order as:

\begin{eqnarray}\label{perturbation}
\textbf{U}'=\emph{M}\cdot \textbf{U}
\end{eqnarray}

\noindent where the prime denotes the derivative with respect to
$N$. The coefficients of the perturbation equations form a
$4\times4$ matrix $\emph{M}$ whose eigenvalues determine the type
and stability of the critical points. The only physically
meaningful critical point corresponding to the autonomous system
(\ref{auto}) is $(x,y,z,\xi)=(-\frac{\lambda}{\sqrt{6}},
\sqrt{1+\frac{\lambda^2}{6}}, 0, 0)$, which corresponds to the
eigenvalues $(0, -3-\frac{\lambda^2}{2}, -3\gamma-\lambda^2,
-3+\frac{\lambda^2}{2})$. Therefore, it is a stable node of the
autonomous system when $\lambda^2<6$. This corresponds to a late
time attractor solution which is a phantom dominated epoch
$\Omega_R=1$ and an equation of state $w=-\frac{\lambda^2}{3}-1$.
Unlike the singlet phantom field in exponential potential
\cite{hao}, the introduction of internal symmetry imposes a lower
bound to the parameter $\lambda^2$ for attractor solution, which
equivalent to the equation of state $w>-3$. This is a very
interesting characteristic of the O(\textit{N}) phantom field. In
Fig. 1 and Fig. 2, we show the numerical results of the phase
space of the system. In this plots, we choose the parameter
$\gamma=1$ and $\lambda=1.2$.
\begin{figure}
\epsfig{file=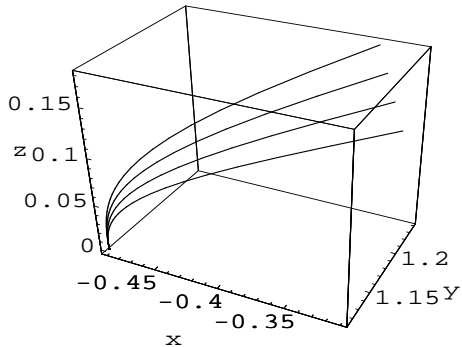,height=3.3in,width=4.0in} \caption{The phase
diagram of the O(\emph{N}) Phantom system in terms of x, y, z for
different initial x, y, z and $\xi$.}
\end{figure}
\begin{figure}
\epsfig{file=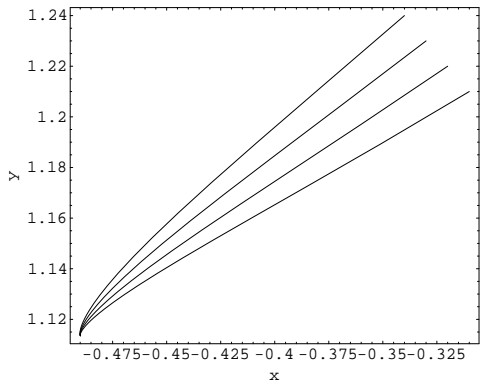,height=2in,width=2.5in} \caption{The
projection of the phase space in x y plane.}
\end{figure}
\begin{figure}
\epsfig{file=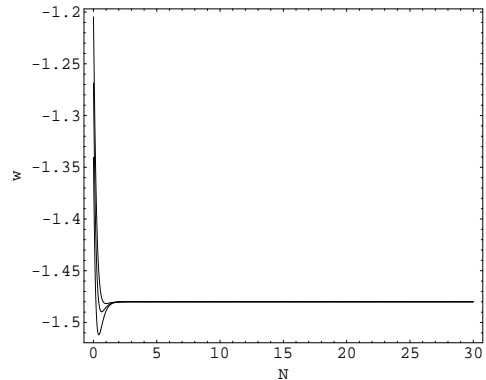,height=2in,width=2.5in} \caption{The
evolution of the equation of state $w$ vs. $N$ for different
initial x, y, z and $\xi$.}
\end{figure}

\section{Reconstruction and discussion}

Now, we will correlate the potential with red shift. To do so,
following the earlier study in this field\cite{turner2}, we
introduce the quantity
\begin{equation}
  r(z)=\int^{t_0}_{t(z)}\frac{du}{a(u)}=\int^z_0\frac{dx}{H(x)}
\end{equation}

\noindent which is the Robertson-Walker coordinate distance to an
object at red-shift $z$. Also, we denote
\begin{equation}\label{rhom}
\rho_M=\frac{3\Omega_M H_0^2(1+z)^3}{8\pi G}
\end{equation}

\noindent where $H_0$ is the present Hubble constant, $\Omega_M$
is the fraction of non-relativistic matter to the critical
density. We then readily have
\begin{eqnarray}
  (\frac{\dot{a}}{a})^2&=&H(z)^2=\frac{1}{(dr/dz)^2}\\\nonumber
 \frac{\ddot{a}}{a}&=&\frac{1}{(dr/dz)^2}+(1+z)\frac{d^2r/dz^2}{(dr/dz)^3}\\\nonumber
  \frac{dz}{dt}&=&-(1+z)H(z)=-(1+z) \frac{dr}{dz}
\end{eqnarray}

\noindent From the above equations, it is not difficult to derive
the reconstruction equations as
\begin{eqnarray}\label{vrecons}
V[R(z)]=&&\frac{1}{8\pi
G}\Big[\frac{3}{(dr/dz)^2}+(1+z)\frac{d^2r/dz^2}{(dr/dz)^3}
\Big]\\\nonumber&&-\frac{3\Omega_M H_0^2(1+z)^3}{16\pi G}
\end{eqnarray}

\begin{eqnarray}\label{Rrecons}
\Big(\frac{dR}{dz}\Big)^2+&&\frac{\Omega^2}{R^2}(1+z)^4\Big(\frac{dr}{dz}
\Big)^2\\\nonumber&&=\frac{(dr/dz)^2}{4\pi
G(1+z)^2}\Big[\frac{(1+z)(d^2/dz^2)}{(dr/dz)^3}+\frac{3}{2}(1+z)^3
\Big]
\end{eqnarray}
\noindent Eq.(\ref{vrecons}) is the same as those ordinary
quintessence while the Eq.(\ref{Rrecons}) is different in that
there is a sign difference: the right hand side of the
Eq.(\ref{Rrecons}) is positive while it is negative in
conventional O(\textit{N}) quintessence model.

Up to now, the observation data do not tell us what should be the
nature of dark energy. But the future observation will be helpful
to determine whether the dark energy is phantom, quintessence, or
cosmological constant. If the equation of state $w<-1$ is
completely confirmed by observations, then its implications to
fundamental physics would be astounding, since it cannot be
achieved with substance with canonical lagrangian. Phantom field
could be a good candidate for such substance because of its
negative kinetic energy and simplicity. In this paper, we study
 a class of new phantom models, in which the scalar field possesses a O(\textit{N}) internal
 symmetry. In a specific model, in
which the potential of the phantom fields is an exponential
potential, we show that there exists an attractor solution when
$\lambda^2<6$ and the attractor solution corresponds to a phantom
energy dominant phase and an equation of state
$w=-\frac{\lambda^2}{3}-1$. Most strikingly, the introduction of
O(\textit{N}) symmetry imposes a restriction to the existence of
the attractor solution $\lambda^2<6$, which accordingly puts a
lower bound to the equation of state $w>-3$. On the other hand,
the speed of sound of the dark energy is defined by
$c_s^2=\frac{p_{,X}}{\rho_{,X}}=\frac{L_{\phi,X}}{L_{\phi,X}+2XL_{\phi,X
X}}$, where $p=L_{\phi}(\phi, X)$ and
$\rho=2XL_{\phi,X}-L_{\phi}(\phi, X)$ with
$X=\frac{1}{2}(\partial_{\mu}\phi)^2$. So, The model in this paper
will produce an acceptable sound of speed $c_s^2=1$.

\acknowledgments{ This work was partially supported by NKBRSF
under Grant No. 1999075406}

\end{document}